\documentclass[natbib]{svjour3}                     
\smartqed  
\usepackage{graphicx}
\usepackage{epsfig,verbatim}
\usepackage{amssymb}
\usepackage{enumerate}
\usepackage{mathptmx}      
\journalname{Space Science Reviews}
\begin{document}

\title{The Pioneer Anomaly in the Light of New Data
}


\author{Slava G. Turyshev        \and
        Viktor T. Toth 
}


\institute{Slava G. Turyshev \at
           Jet Propulsion Laboratory,
       California Institute of Technology,\\
       4800 Oak Grove Drive, Pasadena, CA 91109 USA \\
              \email{turyshev@jpl.nasa.gov}           
           \and
           Viktor T. Toth \at
           Ottawa, ON  K1N 9H5, Canada\\
              \email{vttoth@vttoth.com}
}

\date{Received: date / Accepted: date}

\maketitle

\begin{abstract}
The radio-metric tracking data received from the Pioneer 10 and 11 spacecraft from the distances between 20--70 astronomical units from the Sun has consistently indicated the presence of a small, anomalous, blue-shifted Doppler frequency drift that limited the accuracy of the orbit reconstruction for these vehicles.  This drift was interpreted as a sunward acceleration of $a_P = (8.74\pm 1.33)\times 10^{-10}$~m/s$^2$ for each particular spacecraft. This signal has become known as the Pioneer anomaly; the nature of this anomaly is still being investigated.

Recently new Pioneer 10 and 11 radio-metric Doppler and flight telemetry data became available.  The newly available Doppler data set is much larger when compared to the data used in previous investigations and is the primary source for new investigation of the anomaly.  In addition, the flight telemetry files, original project documentation, and newly developed software tools are now used to reconstruct the engineering history of spacecraft. With the help of this information, a thermal model of the Pioneers was developed to study possible contribution of thermal recoil force acting on the spacecraft.  The goal of the ongoing efforts is to evaluate the effect of on-board systems on the spacecrafts' trajectories and possibly identify the nature of this anomaly.

Techniques developed for the investigation of the Pioneer anomaly are applicable to the New Horizons mission. Analysis shows that anisotropic thermal radiation from on-board sources will accelerate this spacecraft by $\sim41\times 10^{-10}\,\mathrm{m}/\mathrm{s}^2$. We discuss the lessons learned from the study of the Pioneer anomaly for the New Horizons spacecraft.

\end{abstract}

\section{Introduction}
\label{sec:intro}

The first spacecraft to leave the inner solar system \citep{JPL1998,JPL1999,JPL2002,JPL2005}, Pioneer 10 and 11 were designed to conduct an exploration of the interplanetary medium beyond the orbit of Mars and perform close-up observations of Jupiter during the 1972-73 Jovian opportunities.

The spacecraft were launched in March 1972 (Pioneer 10) and April 1973 (Pioneer 11) on top of identical three-stage Atlas-Centaur launch vehicles. After passing through the asteroid belt, Pioneer 10 reached Jupiter in December 1973. The trajectory of its sister craft, Pioneer 11, in addition to visiting Jupiter in 1974, also included an encounter with Saturn in 1979 \citep{JPL2002,MDR2005}.

After the planetary encounters and successful completion of their primary missions, both Pioneers continued to explore the outer solar system. Due to their excellent health and navigational capabilities, the Pioneers were used to search for trans-Neptunian objects and to establish limits on the presence of low-frequency gravitational radiation \citep{PC202}.

Eventually, Pioneer 10 became the first man-made object to leave the solar system, with its official mission ending in March 1997. Since then, NASA's Deep Space Network (DSN) made occasional contact with the spacecraft. The last successful communication from Pioneer 10 was received by the DSN on 27 April 2002. Pioneer 11 sent its last coherent Doppler data in October 1990; the last scientific observations were returned by Pioneer 11 in September 1995.

The orbits of Pioneer 10 and 11 were reconstructed primarily on the basis of radio-metric Doppler tracking data. The reconstruction between heliocentric distances of 20--70 AU yielded a persistent small discrepancy between observed and computed values \citep{JPL1998,JPL2002,JPL1999,JPL2005}. After accounting for known systematic effects, the unmodeled change in the Doppler residual for Pioneer 10 and 11 is equivalent to an approximately sunward constant acceleration of
\begin{equation}
a_P = (8.74\pm 1.33)\times 10^{-10}~\mathrm{m/s}^2.
\end{equation}

The magnitude of this effect, measured between heliocentric distances of 40--70~AU, remains approximately constant within the 3~dB gain bandwidth of the high-gain antenna \citep{Turyshev:2005zm,MDR2005}.  The nature of this anomalous acceleration remains unexplained; this signal has become known as the Pioneer anomaly.

There were numerous attempts in recent years to provide an explanation for the anomalous acceleration of Pioneer 10 and 11. These can be broadly categorized as either invoking conventional mechanisms or utilizing principles of ``new physics''.

Initial efforts to explain the Pioneer anomaly focused on the possibility of on-board systematic forces. While these cannot be conclusively excluded \citep{JPL2002,JPL2005}, the evidence to date did not support these mechanisms: it was found that the magnitude of the anomaly exceeds the acceleration that these mechanisms would likely produce, and the temporal evolution of the anomaly differs from that which one would expect, for instance, if the anomaly were due to thermal radiation of a decaying nuclear power source.

Conventional mechanisms external to the spacecraft were also considered. First among these was the possibility that the anomaly may be due to perturbations of the spacecrafts' orbits by as yet unknown mass distributions in the Kuiper belt. Another possibility is that dust in the solar system may exert a drag force, or it may cause a frequency shift, proportional to distance, in the radio signal. These proposals could not produce a model that is consistent with the known properties of the Pioneer anomaly, and may also be in contradiction with the known properties of planetary orbits.

The value of the Pioneer anomaly happens to be approximately $cH_0$, where $c$ is the speed of light and $H_0$ is the Hubble constant at the present epoch. Attempts were made to exploit this numerical coincidence to provide a cosmological explanation for the anomaly, but it has been demonstrated that this approach would likely produce an effect with the opposite sign \citep{JPL2002,MDR2005}.

As the search for a conventional explanation for the anomaly appeared unsuccessful, this provided a motivation to seek an explanation in ``new physics''. No such attempt to date produced a clearly viable mechanism for the anomaly \citep{MDR2005}.

Here we report on the status of the recovery of the Pioneers' radiometric Doppler data and flight telemetry and their usefulness for the analysis of the Pioneer anomaly.

\section{New Doppler data and their preliminary analysis}
\label{sec:doppler}

The inability to explain the anomalous behavior of the Pioneers with conventional physics has resulted in a growing discussion about the origin of the detected signal. The limited size of the previously analyzed data set also limits our current knowledge of the anomaly. To determine the origin of $a_P$ and especially before any serious discussion of new physics can take place, one must analyze the entire set of radio-metric Doppler data received from the Pioneers.

Since 2002, multiple on-going efforts have contributed significantly to our ability to explore and comprehend the nature of the Pioneer anomaly \citep{Turyshev:2005zm,MDR2005,Turyshev:2005vj,MDR2006,MDR2008}. Most notable among these are: (i) the availability of an extended Doppler data set; (ii) the recovery of spacecraft telemetry; (iii) the recovery of Pioneer project documentation; (iv) the development of a comprehensive thermal model of the spacecraft; (v) the development of new methods to incorporate thermal telemetry into orbit determination; and (vi) several independent confirmations of the Pioneer anomaly.
These developments led to the formulation of a comprehensive strategy to establish reliably the temporal dependence and direction of the anomalous acceleration, and correlate it with revised estimates of the thermal recoil force.

\subsection{The extended Pioneer Doppler data set}
\label{sec:pio-doppler}

Immediately after the results of the first major study of the anomaly  were announced \citep{JPL2002}, a focused effort began at JPL to recover as much Doppler data as possible. Initially, it was hoped that nearly all the Doppler record of the Pioneer 10 and 11 spacecraft from 1972 until the end of their respective missions can be recovered. Unfortunately, this proved much more difficult than anyone anticipated at the time.

Recovery of radio-metric data for a mission operating for more then 30 years is an effort that was never attempted before. Indeed, 30 years is a long time, presenting many unique challenges, including changes in the data formats, navigational software, as well as supporting hardware \citep{MDR2005}. Even the DSN configuration had changes -- new stations were built, and some stations were moved, upgraded and reassigned. By 2005 all the DSN data formats, navigational software used to support Pioneers, all the hardware used to read, write and maintain the data have become obsolete and are no longer operationally supported by existing NASA protocols. The main asset of the entire mission support -- its people -- changed the most, as personnel with the necessary expertise to answer questions or shed light on obscure details are either retired or no longer with us.

Despite these multiple complexities \citep{MDR2005,MDR2006,MDR2008}, the transfer of the available Pioneer Doppler data to modern media formats has been completed. However, as a result of these issues, less data is available for analysis than what was initially hoped. Nonetheless, we now have a significantly expanded data set that is available for study.

For Pioneer 10, good quality Doppler data are available covering the period between 1980 and 1995. Some additional data from the year 2000 also appear usable. Yet more data files, from 1996-97, may be usable if absent ramp information can be recovered from the transmitting DSN stations.

For Pioneer 11, good quality Doppler data are available from the periods 1983-85, and 1987-90. Data from 1980-82 and from 1986, while recovered, appear unusable.

Additional data covering the Jupiter encounter of Pioneer 10, and the Saturn encounter of Pioneer 11, are also available and appear to be of good quality. Modeling encounters with gas giants is fraught with additional difficulties, as both gravitational and nongravitational effects of the complex planetary environment must be modeled with precision. On the other hand, the rapid changes in the spacecrafts' velocities can help uncover small systematic effects that are otherwise difficult to observe.

\subsection{A strategy to find the origin of the Pioneer anomaly}
\label{sec:objectives}

The primary objective of the new investigation is to determine the origin of the Pioneer anomaly. Specifically, the investigation intends to accomplish the following three major objectives:

\begin{itemize}
\item[I.] By using the early mission Doppler data we aim: (i) to determine the true direction of anomalous acceleration by discriminating between the four possible directions: sunward, Earth-pointing, along the velocity vector, or along the spin-axis, and (ii) to study the physics of the planetary encounters and to learn more about the onset of the anomaly (see Figs.~6-7 in \citep{JPL2002});
\item[II.] Analysis of the entire set of Doppler data: (iii) to study the temporal behavior of the signal, and (iv) to perform a comparative analysis of the individual anomalous accelerations of the two Pioneers with data taken from similar heliocentric distances; and finally
\item[III.] The newly recovered telemetry information from the spacecrafts' thermal, electrical, power, propulsion, and communication subsystems will be used (v) to investigate contributions of known sources of on-board systematics and study their effects on $a_P$. This investigation of the on-board small forces will be done in conjunction with the analysis of the Doppler data, thereby strengthening the ultimate outcome.
\end{itemize}

The new extended data set enables us to investigate the Pioneer anomaly with the entire available Pioneer 10 and 11 radiometric Doppler data \citep{MDR2005,MDR2008}. In particular, it is used to build a thermal/electrical/dynamical model of the Pioneer vehicles and verify it with the actual data from the telemetry; the goal here is the development of a model that can be used to calibrate the Doppler anomaly with respect to the on-board sources of dynamical noise.

\subsubsection{Direction of the Pioneer anomaly}

Analysis of the earlier data is critical in establishing a precise time history of the effect. With the radiation pattern of the Pioneer antennae and the lack of precise navigation in the plane of the sky, the determination of the exact direction of the anomaly was a difficult task \citep{JPL2002}. While in deep space, for standard antennae without good 3-D navigation,  the directions: (i) towards the Sun, (ii) towards the Earth, (iii) along the direction of motion of the craft, or (iv) along the spin axis, are all observationally synonymous \citep{Nieto:2003rq}.

The four possible directions for the anomaly all indicate a different physical mechanism.  Specifically, if the anomaly is: (i) in the direction {\it towards the Sun}, this would indicate a force originating from the Sun, likely signifying a need for gravity modification; (ii) in the direction {\it towards the Earth}, this would indicate an effect on signal propagation or a time signal anomaly impacting the design of the DSN hardware and space flight-control methods; (iii) in the direction {\it of the velocity vector}, this would indicate an inertial force or a drag force providing support for a media-dependent origin; or, finally, (iv) in {\it the spin-axis direction}, this would indicate an on-board systematic, which is the most plausible explanation for the effect.  The corresponding directional signatures of these four directions are distinct and could be extracted from the data \citep{Nieto:2003rq,Turyshev:2005zm,MDR2006}.

The increased data span, and especially the earlier data segment, is crucial in our ability to determine the direction of the signal and its true nature. If the anomaly is due to on-board effects, direct along the spin axis, the solution for the anomaly will be different after each re-pointing of the spacecraft, namely, every re-pointing will produce a step-function discontinuity in the solution for the $a_P$.  Earlier in the mission, there were many of these re-pointing maneuvers; understanding their impact on the anomaly would be a very important activity of the proposed analysis of the earlier data \citep{Turyshev:2005zm}.

With the early data, we expect to improve the sensitivity of the solutions in the directions perpendicular to the line-of-sight by at least an order of magnitude. We started to analyze the early parts of the trajectories of the Pioneers with the goal of determining the true direction of the Pioneer anomaly and possibly its origin \citep{Nieto:2003rq,Turyshev:2005zm,Turyshev:2005vj}; results will be reported elsewhere.

\subsubsection{Study of the planetary encounters}

It is alarming is that the early Pioneer 10 and 11 data (before 1987) was never analyzed in detail, especially in regards to the effect of on-board systematics.  For instance, a nearly constant anomalous acceleration seems to exist in the data of Pioneer 10 as close as 27 AU from the Sun \citep{JPL2002,JPL2005}.  Pioneer 11, beginning just after Jupiter flyby, shows a small value for the anomaly during the Jupiter-Saturn cruise phase in the interior of the solar system.  However, right at Saturn encounter, when the craft passed into a hyperbolic escape orbit, analysis of early navigation results tentatively shows a fast increase in the anomaly to its canonical value \citep{Nieto:2003rq,Turyshev:2005vj}.

We first study the Saturn encounter for Pioneer 11.  We will use the data for nearly two years surrounding this event.  If successful, we should be able to learn the mechanism that led to the onset of the anomaly during the flyby. The Jovian encounters are also of interest; however, they were in the region much too close to the Sun.  One expects large contributions from the standard sources of acceleration noise that exist at heliocentric distances $\sim5$ AU.  We use a similar strategy as with the Saturn encounter and will attempt to make full use of the data available.

\subsubsection{Study of the temporal evolution of the anomaly}

The same investigation as above is used to revisit the question of collimated thermal emission by studying the temporal evolution of the anomaly. If the anomaly is due to the on-board nuclear fuel inventory ($^{238}$Pu) present on the vehicles and the related heat recoil force, one expects that a decrease in the anomaly's magnitude will be correlated with $^{238}$Pu decay with a half-life of 87.74 years. The previous analysis of 11.5 years of data \citep{JPL2002} found no support for a thermal mechanism. However, \cite{Markwardt:2002ma} and \cite{Toth:2009hx} did not rule out this possibility finding an appropriate trend in the solution for $a_P$.  The now available 30-year interval of data (20 years for Pioneer 11) may demonstrate the effect of a $\sim21$~\% reduction in the heat contribution to the anomaly.  This behavior would strongly support a thermal origin of the effect \citep{MDR2008}.

Asymmetrically radiated heat due to available on-board thermal inventory, if appropriately directed, may result in a recoil force with properties similar to those of the Pioneer anomaly.  This is the ``primary suspect'' for the origin of the effect.  Therefore, an investigation of heat exchange mechanisms that may lead to a thermal thrust on the craft is an important part our effort. This investigation is done in two ways: (i) to use the longest possible set of Doppler data to study the temporal evolution of $a_P$ and (ii) to build a model of thermal thrust of the Pioneer craft and use available telemetry data to derive the resulted acceleration. If the anomaly is due to a thermal mechanism, the two methods should agree.

The much extended data span, augmented by all the ancillary spacecraft information, will help to study a signature of the exponential decay of the on-board power source \citep{Markwardt:2002ma,Toth:2009hx}.
The wealth of the recently acquired data presents an exciting opportunity to learn more about the anomaly in various regimes and will help to determine the nature of this anomalous signal.

\subsubsection{Analysis of the individual trajectories for both Pioneers}

The much larger newly recovered sets of the Pioneer 10 and 11 Doppler data make it possible to study the properties of the individual acceleration solutions for both Pioneers obtained with the data collected from similar heliocentric distances.  The limited data set used in the previous analysis \citep{JPL1998,JPL2002}  precluded such a comparison; however, the now available data allows such an investigation.

Previously, even though we had individual solutions from the two craft, the fact remained that $a_{P10}$ and $a_{P11}$ were obtained from data segments that not only were very different in length (11.5 and 3.75 years), but they were also taken from different heliocentric distances. We anticipate that analysis of their data from similar heliocentric regions will help us better understand the properties of the anomaly, especially if it were to be attributed to an on-board systematic source. Analysis of the individual data would also help to calibrate the final solution for the anomaly by properly accounting for the individual properties of the spacecraft.

\subsubsection{Investigation of on-board systematics}

The availability of telemetry information makes it possible to conduct a detailed investigation of the on-board systematic forces as a source of the anomalous acceleration. Here we consider forces that are generated by on-board spacecraft systems and that are thought to contribute to the constant acceleration seen in the analysis of the Pioneer Doppler data (see Section~\ref{sec:system-mdr} for discussion).

\section{Using flight telemetry to study the spacecrafts' behavior}
\label{sec:system-mdr}

All transmissions of both Pioneer spacecraft, including all engineering telemetry, were archived \citep{MDR2005} in the form of files containing Master Data Records (MDRs). Originally, MDRs were scheduled for limited retention. Fortunately, the Pioneers' mission records avoided this fate: with the exception of a few gaps in the data \citep{MDR2005} the entire mission record has been saved. These recently recovered telemetry readings are important in reconstructing a complete history of the thermal, electrical, and propulsion systems for both spacecraft. This, may data lead to a better determination of the crafts' acceleration due to on-board systematic effects.

\subsection{The Pioneer spacecraft}

\begin{figure}
\includegraphics[width=\linewidth]{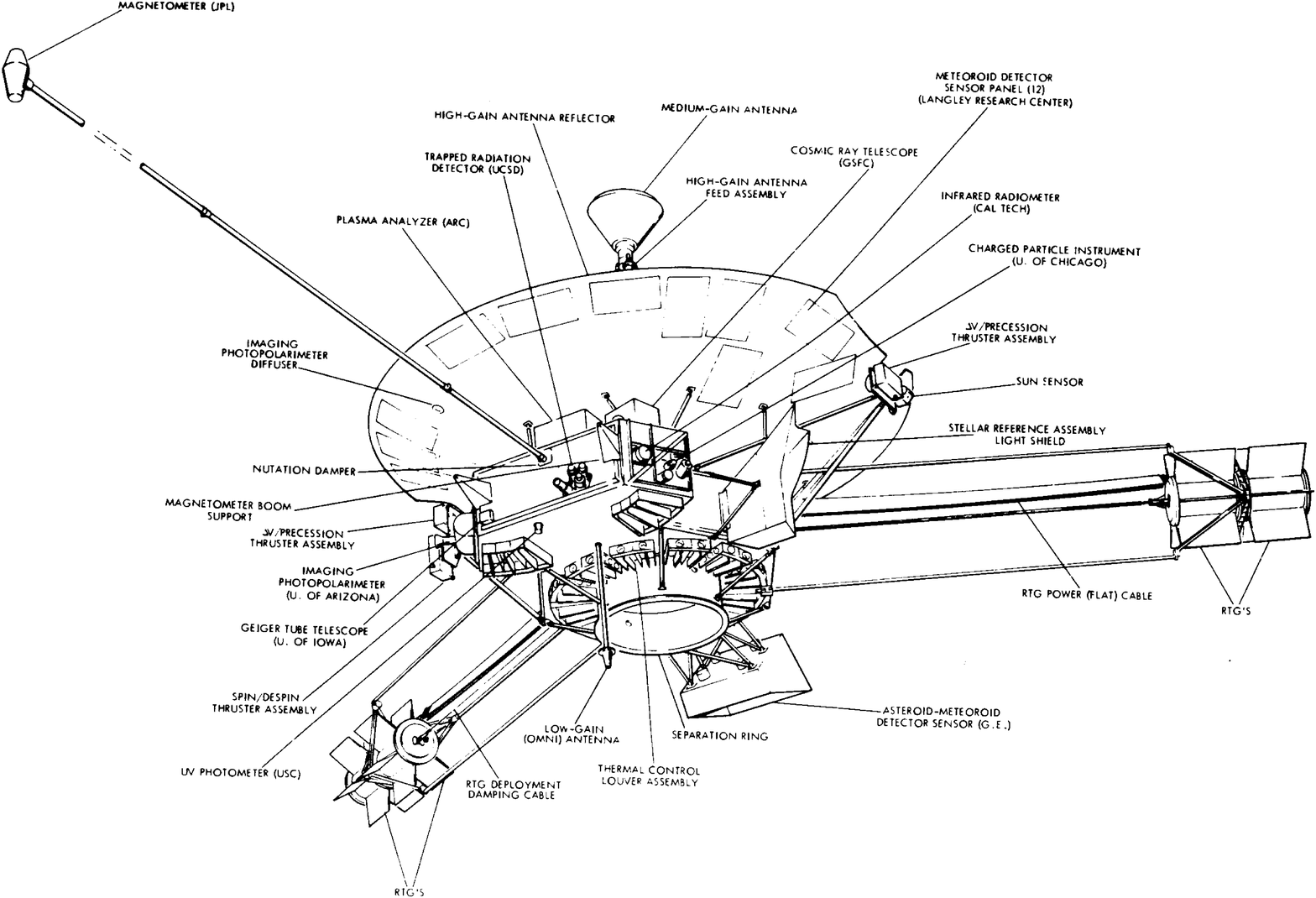}
\caption{A drawing of the Pioneer spacecraft.}
\label{fig:pioneer}
\end{figure}

As evident from Fig.~\ref{fig:pioneer}, the appearance of the Pioneer spacecraft is dominated by the 2.74~m diameter high gain antenna (HGA). The spacecraft body, located behind the HGA, consists of a larger, regular hexagonal compartment housing the propellant tank and spacecraft electronics; an adjacent, smaller compartment housed science instruments. The spacecraft body is covered by multilayer thermal insulating blankets, except for a louver system located on the side opposite the HGA, which was activated by bimetallic springs to expel excess heat from the spacecraft.

Each spacecraft was powered by four radioisotope thermoelectric generators (RTGs) mounted in pairs at the end of two booms, approximately three meters in length, extended from two sides of the spacecraft body at an angle of 120$^\circ$. A third boom, approximately 6 m long, held a magnetometer.

The total (design) mass of the spacecraft was $\sim$250~kg at launch, of which 27~kg was propellant \citep{PC202}.

For the purposes of attitude control, the spacecraft were designed to spin at the nominal rate of 4.8~rpm. Six small monopropellant (hydrazine) thrusters, mounted in three thruster cluster assemblies, were used for spin correction, attitude control, and trajectory correction maneuvers (see Fig.~\ref{fig:elec}).

\subsection{Compartment temperatures and thermal radiation}

If the Pioneer anomaly is due to anisotropically emitted thermal radiation, one expects to find a near constant supply of heat radiated off the back of the spacecraft that would produce a thermal recoil force with the well established properties \citep{JPL2005}.
The lack of constancy of heat dissipated during the longest Doppler segment analyses (i.e. 11.5 years of the Pioneer 10 data \citep{JPL2002}) appears to have invalidated the hypothesis.

The newly acquired data (both Doppler and telemetry) is very valuable for the investigation as it contributes to addressing this possibility. We also have a much larger segment of Doppler data that will be used to analyze these heat dissipation processes on the vehicles. In addition, we now have the actual design, fabrication, testing, pre- and in-flight calibration data that characterize the  Pioneer craft performance for the duration of their missions. Finally, we have all the detailed information on properties of the spacecraft and the data needed to reconstruct the behavior of its major components, including electrical power and thermal subsystems.

This data tells precisely at what time the louvers were open and closed, when a certain instrument was powered ``on'' and ``off'', what was the performance of the battery, shunt current and all electric parts of the spacecraft. This information is being used in the  development of a model of the Pioneers that is needed to establish the true thermal and electrical power dissipation history of the vehicles and also to correlate major events on the Pioneers (such as powering ``on'' or ``off'' certain instruments or performing a maneuver) with the anaysis of the available Doppler data.

\subsection{Telemetry overview}

Telemetry formats can be broadly categorized as science formats versus engineering formats. Telemetry words included both analog and digital values. Digital values were used to represent sensor states, switch states, counters, timers, and logic states. Analog readings, from sensors measuring temperatures, voltages, currents and more, were encoded using 6-bit words. This necessarily limited the sensor resolution and introduced a significant amount of quantization noise. Furthermore, the analog-to-digital conversion was not necessarily linear; prior to launch, analog sensors were calibrated using a fifth-order polynomial. Calibration ranges were also established; outside these ranges, the calibration polynomials are known to yield nonsensical results.

With the help of the information contained in these words, it is possible to reconstruct the history of RTG temperatures and power, radio beam power, electrically generated heat inside the spacecraft, spacecraft temperatures, and propulsion system history.

Relevant on-board telemetry falls into two categories: temperature and electrical measurements. In the first category, we have data from several temperature sensors on-board, most notably the fin root temperature readings for all four RTGs. Figure~\ref{fig:tempsens} shows the location of most temperature sensors on board for which readings are available. Other temperature sensors are located at the RTGs and inside the propellant tank.

The electrical power profile of the spacecraft can be reconstructed to a reasonable degree of accuracy using electrical telemetry measurements. Available are the individual voltage and current readings for the RTGs, the readings on the main bus voltage and current, as well as the shunt current; known are the power on/off state of most spacecraft subsystems. From these and other readings, one can calculate the complete electrical profile of the spacecraft.

\subsection{RTG temperatures and anisotropic heat reflection}

It has been argued that the anomalous acceleration may be due to anisotropic reflection of the heat coming from the RTGs off the back of the spacecraft high gain antennae. Note that only $\sim$65~W of directed constant heat is required to explain the anomaly, which certainly is not a great deal of power when the craft has heat sources capable of producing almost 2.5~kW of heat at the beginning of the missions. However, using available information on the spacecraft and RTG designs, \cite{JPL2002} estimated that only 4\,W of directed power could be produced by this mechanism. Adding an uncertainty of the same size, they estimated a contribution to the anomalous acceleration from heat reflection to be $a_{\rm hr}= (-0.55 \pm 0.55) \times 10^{-10}\,\mbox{m/s}^2.$ Furthermore, if this mechanism were the cause of the anomaly, ultimately an unambiguous decrease in the size of $a_P$ should be observed, because the RTGs' radioactively produced radiant heat is decreasing. In fact, one would expect a decrease of about $0.75\times 10^{-10}$\,m/s$^2$ in $a_P$ over the 11.5 year Pioneer 10 data interval if this mechanism were the origin of $a_P$.

Alternative estimates presented put the magnitude of this effect at $\sim$24\,W or the corresponding value for $a_{\rm hr}$ at the level of $a_{\rm hr}\sim -3.3 \times 10^{-10}\,\mbox{m/s}^2$ (see discussion in \citep{Scheffer:2001se}). We comment on the fact that both groups acknowledged that a thermal model for the Pioneer spacecraft is hard to build. However, this seems to be exactly what one would have to do in order to reconcile the differences in analyzing the role of the thermal heat in the formation of the Pioneer anomaly.

It is clear that any thermal explanation should clarify why either the radioactive decay (if the heat is directly from the RTGs \citep{Katz:1998ew}) or electrical power decay (if the heat is from the instrument compartment \citep{Murphy:1998hp}) is not seen. One reason could be that previous analyses used only a limited data set of only 11.5 years when the thermal signature was hard to disentangle from the Doppler residuals or the fact that the actual data on the performance of the thermal and electrical systems was not complete or unavailable at the time the analyses were performed.

The present situation is very different.  Not only do we have a much longer Doppler data segment for both spacecraft, we also have the actual telemetry data on the thermal and electric power subsystems for both Pioneers for the entire lengths of their missions.  The electrical power profile of the spacecraft can be reconstructed to a reasonable degree of accuracy using electrical telemetry measurements.  We have individual voltage and current readings for the RTGs; we have readings on the main bus voltage and current, as well as the shunt current; we know the power on/off state of most spacecraft subsystems.  From these and other readings, the complete electrical profile of the spacecraft can be calculated, as discussed in \citep{MDR2005}.

To utilize this data for the upcoming investigation, it is possible to reconstruct the direction of heat flow (with the help of a thermal model built for this purpose) and study the absorption and re-emission by, and reflection off the craft surfaces \citep{MDR2005,MDR2008}.

In the following discussion, telemetry words are labeled using identifiers in the form of $C_n$, where $n$ is a number indicating the word position in the fixed format telemetry frames.

\begin{figure}
\centering \psfig{file=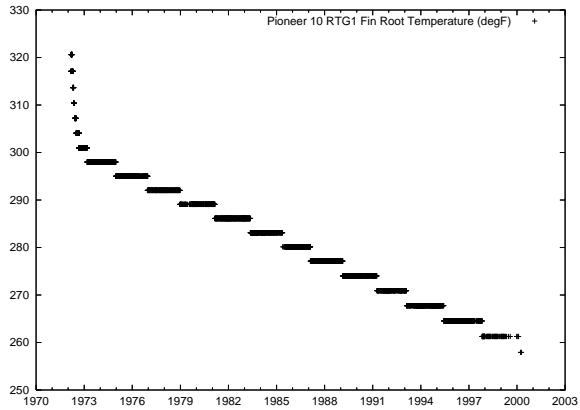, width=0.65\linewidth}
\caption{RTG 1 fin root temperatures (telemetry word $C_{201}$; in $^\circ$F) for Pioneer 10.}
\label{fig:C201}
\end{figure}

The exterior temperatures of the RTGs were measured by one sensor on each of the four RTGs: the so-called ``fin root temperature'' sensor. Telemetry words $C_{201}$ through $C_{204}$ contain the fin root temperature sensor readings for RTGs 1 through 4, respectively. Figure~\ref{fig:C201} depicts the evolution of the RTG 1 fin root temperature for Pioneer 10 (other fin root sensors exhibited similar behavior \citep{MDR2005}).

A best fit analysis confirms that the RTG temperature indeed evolves in a manner consistent with the radioactive decay of the nuclear fuel on board. The results for all the other RTGs on both spacecraft are similar, confirming that the RTGs were performing thermally in accordance with design expectations.

\subsection{RTG power electrically generated heat}

RTG electrical power can be estimated using two sensor readings per RTG, measuring RTG current and voltage. Currents for RTGs 1 through 4 appear as telemetry words $C_{127}$, $C_{105}$, $C_{114}$, and $C_{123}$, respectively; voltages are in telemetry words $C_{110}$, $C_{125}$, $C_{131}$, and $C_{113}$. Combined, these words yield the total amount of electrical power available on board:
{}
\begin{equation}
P_E = C_{110}C_{127} + C_{125}C_{105} + C_{131}C_{114} + C_{113}C_{123}.
\end{equation}

All this electrical power is eventually converted to waste heat by the spacecrafts' instruments, with the exception of power radiated away by transmitters.

\begin{figure}
\centering \psfig{file=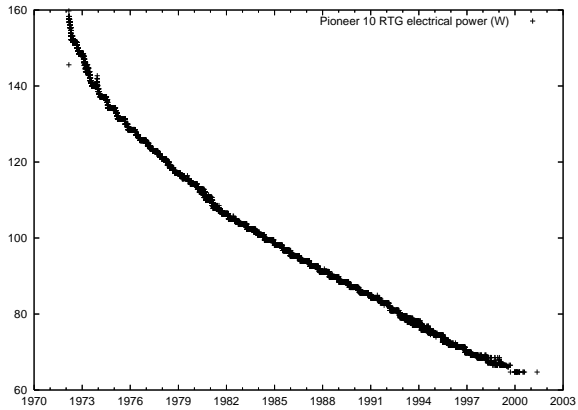, width=0.65\linewidth}
\caption{Changes in total RTG electrical output (in~W) on board Pioneer, as computed using the mission's on-board telemetry.}
\label{fig:elec}
\end{figure}

Whatever remains of electrical energy (Fig.~\ref{fig:elec}) after accounting for the power of the transmitted radio beam is converted to heat on-board. Some of it is converted to heat outside the spacecraft body by externally mounted components.

The Pioneer electrical system is designed to maximize the lifetime of the RTG thermocouples by ensuring that the current draw from the RTGs is always optimal. This means that power supplied by the RTGs may be more than that required for spacecraft operations. Excess electrical energy is absorbed by a shunt circuit that includes an externally mounted radiator plate. Especially early in the mission, when plenty of RTG power was still available, this radiator plate was the most significant component external to the spacecraft body that radiated heat. Telemetry word $C_{122}$ tells us the shunt circuit current, from which the amount of power dissipated by the external radiator can be computed using the known ohmic resistance ($\sim$5.25~$\Omega$) of the radiator plate.

Other externally mounted components that consume electrical power are the Plasma Analyzer ($P_\mathrm{PA} = 4.2$~W, telemetry word $C_{108}$ bit 2), the Cosmic Ray Telescope ($P_\mathrm{CRT} = 2.2$~W, telemetry word $C_{108}$, bit 6), and the Asteroid/Meteoroid Detector ($P_\mathrm{AMD} = 2$~W, telemetry word $C_{124}$, bit 5). Though these instruments' exact power consumption is not telemetered, we know their average power consumption from design documentation, and the telemetry bits tell us when these instruments were powered.

Two additional external loads are the battery heater and the propellant line heaters. These represent a load of $P_\mathrm{LH} = P_\mathrm{BH} = 2$~W (nominal) each. The power state of these loads is not telemetered. According to mission logs, the battery heater was commanded off on both spacecraft on 12 May 1993.

Yet a further external load is the set of cables connecting the RTGs to the inverters. The resistance of these cables is known: it is 0.017~$\Omega$ for the inner RTGs (RTG 3 and 4), and 0.021~$\Omega$ for the outer RTGs (RTG 1 and 2). Using the RTG current readings it is possible to accurately determine the amount of power dissipated by these cables in the form of heat:
{}
\begin{equation}
P_\mathrm{cable} = 0.017(C^2_{114}+C^2_{123}) + 0.021(C^2_{127} + C^2_{105}).
\end{equation}

After accounting for all these external loads, whatever remains of the available electrical power on board is converted to heat inside the spacecraft. So long as the body of the spacecraft is in equilibrium with its surroundings, heat dissipated through its walls has to be equal to the heat generated inside:
{}
\begin{equation}
P_\mathrm{body} = P_E - P_\mathrm{cable} - P_\mathrm{PA} - P_\mathrm{CRT} - P_\mathrm{AMD} - P_\mathrm{LH} - P_\mathrm{BH},
\end{equation}
with all the terms defined above.

\subsection{The thermal control subsystem}

\begin{figure}
\centering \psfig{file=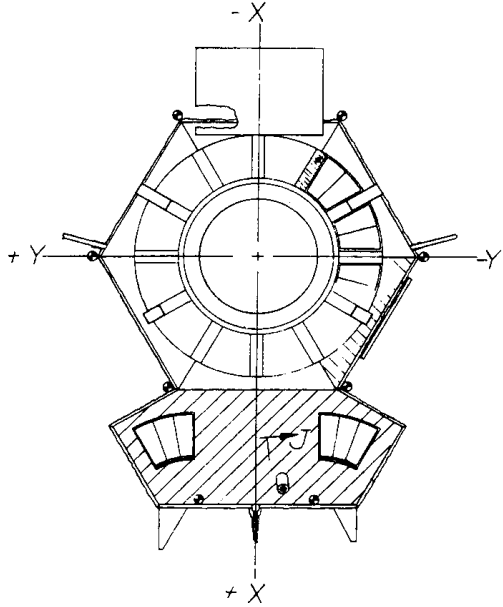, width=0.55\linewidth}
\caption{Bottom view of the Pioneer 10 and 11 vehicle, showing the louver system. A set of 12 2-blade louver assemblies cover the main compartment in a circular pattern; an additional two 3-blade assemblies cover the compartment with science instruments.}
\label{fig:louvers}
\end{figure}

The passive thermal control system consisted of a series of spring-activated louvers (see Fig.~\ref{fig:louvers}). The springs were bimetallic, and thermally (radiatively) coupled to the electronics platform beneath the louvers. The louver blades were highly reflective in the infrared. The assembly was designed so that the louvers fully open when temperatures reach 30$^\circ$C, and fully close when temperatures drop below 5$^\circ$C.

The effective emissivity of the thermal blankets used on the Pioneers is $\epsilon_\mathrm{sides} = 0.085$ \citep{MDR2006}. The total exterior area of the spacecraft body is $A_\mathrm{walls} = 4.92$~m$^2$. The front side of the spacecraft body that faces the HGA has an area of $A_\mathrm{front} = 1.53$~m$^2$, and its effective emissivity, accounting for the fact that most thermal radiation this side emits is reflected by the back of the HGA, can be computed as $\epsilon_\mathrm{front} = 0.0013$. The area covered by louver blades is $A_\mathrm{louv} = 0.29$~m$^2$; the effective emissivity of closed louvers is $\epsilon_\mathrm{louv} = 0.04$ \citep{PC202}. The area that remains, consisting of the sides of the spacecraft and the portion of the rear not covered by louvers is $A_\mathrm{sides}= A_\mathrm{walls} - A_\mathrm{front} - A_\mathrm{louv}$.

Using these numbers, we can estimate the amount of electrically generated heat radiated through the (closed) louver system as a ratio of total electrical heat generated inside the spacecraft body. If we assume that the exterior of the spacecraft is approximately isothermal, we obtain
{}
\begin{equation}
P_\mathrm{louver}=\frac{\epsilon_\mathrm{louv}A_\mathrm{louv}P_\mathrm{body}}
{\epsilon_\mathrm{louv}A_\mathrm{louv}+\epsilon_\mathrm{sides}A_\mathrm{sides}+\epsilon_\mathrm{front}A_\mathrm{front}}=0.042 \,P_\mathrm{body}
\end{equation}

This result is a function of the electrical power generated inside the spacecraft body. However, we also have in our possession thermal vacuum chamber test results of the Pioneer louver system. These results characterize louver thermal emissions as a function of the temperature of the electronics platform beneath the louvers, with separate tests performed for the 2-blade and 3-blade louver assemblies. To utilize these results, we turn our attention to telemetry words representing electronics platform temperatures.

\begin{figure}
\centering
\includegraphics[width=0.70\linewidth]{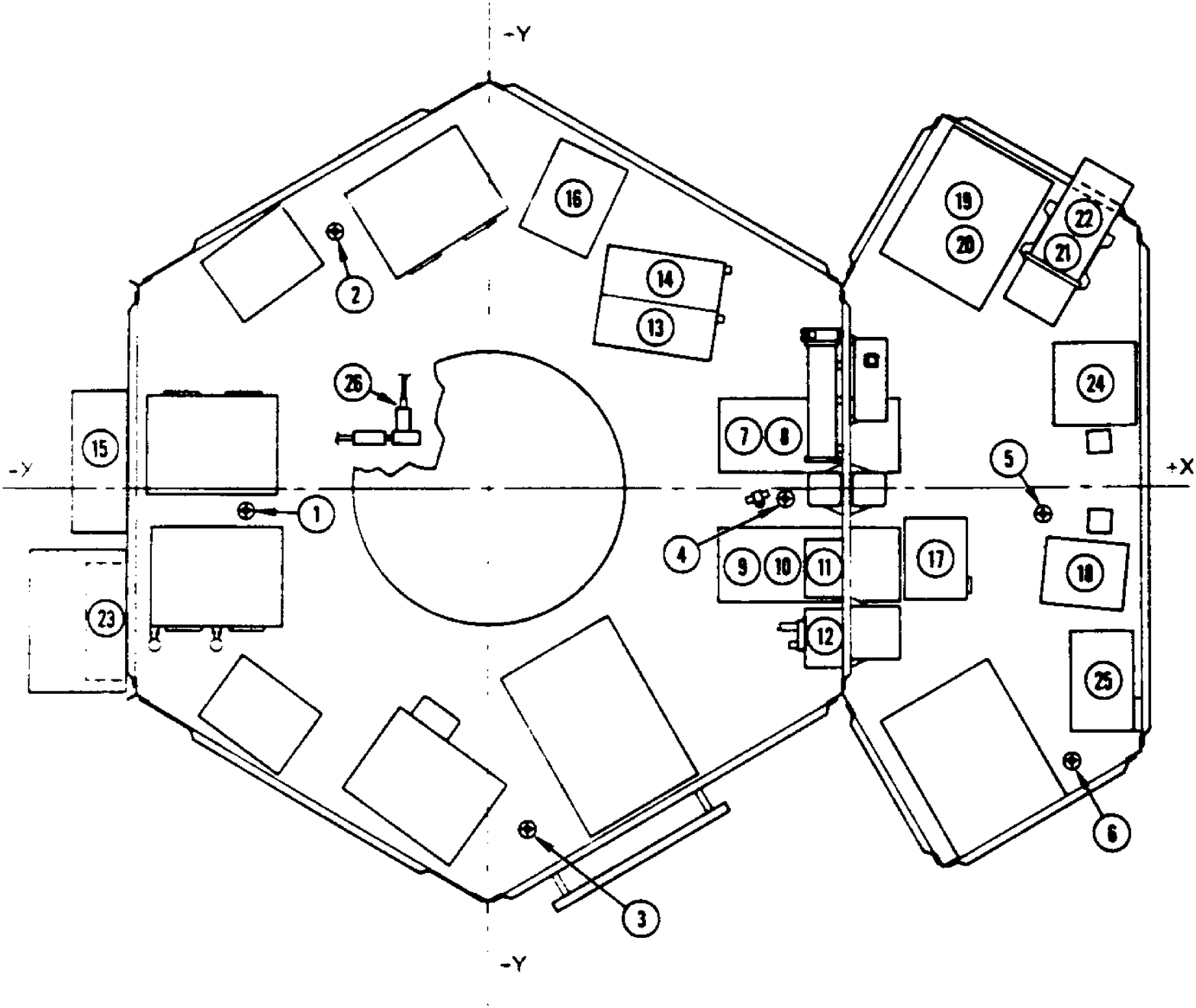}
\caption{Location of thermal sensors in the instrument compartment of Pioneer 10 and 11 \citep{PC202}. Temperature sensors are mounted at locations 1 to 6.}
\label{fig:tempsens}
\end{figure}

There are 6 platform temperature sensors (Fig.~\ref{fig:tempsens}) inside the spacecraft body: 4 are located inside the main compartment, 2 sensors are in the science instrument compartment. The main compartment has a total of 12 2-blade louver blade assemblies; the science compartment has 2 3-blade assemblies.

The thermal vacuum chamber tests provide values for emitted thermal power per louver assembly as a function of the temperature of the electronics platform behind the louver. This allows us to estimate the amount of thermal power leaving the spacecraft body through the louvers, as a function of platform temperatures \citep{MDR2006}, providing means to estimate the amount of heat radiated by the louver system.

The study of the thermal hypothesis as the likely cause of the  Pioneer anomaly is still on-going. Our preliminary thermal modeling indicates that anisotropic thermal radiation may account for some, but not necessarily for all of the anomalous acceleration of the Pioneers.  Rather than treating this result as the ``smoking gun'', we realized that the study of the anomaly requires a thorough and complete understanding and characterization of this thermal recoil force. This means that the most difficult part of our work only just began. Initial results are very intriguing, leading us to believe that we will be able to address all the main objectives of the study of the Pioneer anomaly (see discussion in  Sec.~\ref{sec:objectives}). We will report the results of this investigation elsewhere.

Although significant speculation in the investigation of the Pioneer anomaly has arisen over ``new physics,'' it is likely that the anomaly is systematic in nature.  Still, further high precision tests of this effect might confirm or refute the Pioneer results.   We suggest that the New Horizons spacecraft would provide an opportunity for a test.

\section{Studying the Pioneer anomaly with New Horizons}
\label{sec:nh}

Computation of recoil forces due to thermal momentum transfer is rarely needed in ground-based applications, as the resulting forces are exceedingly small for any practical purpose \citep{Toth:2009se}. However, corresponding effects become important for precision navigated spacecraft such as Pioneer 10 and 11 where, in order to find the origin of the Pioneer anomaly, one needs to account for all minuscule dynamical disturbances of the spacecraft trajectory and its attitude at the level of $\sim10^{-10}$~m/s$^2$ \citep{JPL2002,JPL2005}. Reaching this level of accuracy depends on i) thermal spacecraft design that is free from introducing parasitic sources of non-gravitational noise and ii) an ability to compensate for unwanted forces during orbit determination process. One particular example to apply our methods is the New Horizons mission to Pluto and beyond.

New Horizons began its decade long journey to Pluto and beyond on January 19, 2006. The spacecraft and its mission bear many similarities to the Pioneer 10 and 11 spacecraft and missions. Both types of spacecraft are spin-stabilized. They are powered by RTGs. Both spacecraft utilize a louver system to vent excess heat from the spacecraft body. Their appearance is dominated by a large HGA with the spacecraft body located behind it. They all follow hyperbolic escape trajectories. Due to these similarities, the possibility exists that New Horizons might provide a means to further investigate the Pioneer anomaly \citep{JPL2002}.

There are some differences between the two spacecraft. New Horizons uses X-band frequencies for communication, which allow for better quality radiometric data. Also, during the New Horizons cruise phase to Pluto, the spacecraft is in ``hibernation'' mode, with low operational activity and minimal tracking.

The study of the anomalous acceleration of the Pioneer 10 and 11 spacecraft \citep{JPL2002,MDR2006} demonstrates the importance of taking into account very small forces that might affect the accuracy of spacecraft navigation. We developed an approach to investigate this claim, thereby establishing a foundation for the study of the Pioneer anomaly using recently recovered extended set of radio-metric Doppler and flight telemetry data \citep{MDR2006,MDR2008}.

During the course of our work, we learned some lessons that are directly applicable to New Horizons. Specifically, we found that the two main sources of thermal acceleration are asymmetrically reflected heat from the RTGs, and electrical heat generated inside the spacecraft's body.

Based solely on publicly available information sources, we can develop a preliminary estimate of these heat sources on the trajectory of New Horizons. We start with electrically generated heat.

\subsection{Electrically generated heat}

The New Horizons power source is a GPHS-RTG fueled with 61 fuel elements. The nominal electrical power of a GPHS-RTG fueled with the maximum of 72 fuel elements is 285~W, and its total power is $\sim$4.3~kW.
Thus, we calculate the electrical power of the New Horizons RTG at the time of launch as $P_E=240$~W, and its total power, 3.36~kW. These values agree with values published in  \citep{NH2006}.

To maximize the lifetime of the thermocouple elements inside the RTG, the spacecraft is designed to always draw the optimal amount of current from the RTG; excess electrical power is directed to a shunt circuit. Shunted power may be dissipated in the form of heat either inside the spacecraft body or outside, depending on thermal conditions. This design suggests that the amount of electrical power converted to heat inside the craft body remains approximately constant throughout the mission, at least after it leaves the inner solar system and is no longer significantly heated by the Sun.

Accordingly, we assume that the amount of electrically generated heat inside the spacecraft body will be the nominal figure of 150~W,
constant throughout the mission except for the very early phases, when solar heating is significant, and the very late phases, when RTG power drops to a low level.

Heat will leave the interior of the spacecraft body through its walls. The walls are generally covered with multilayer insulation, whose typical effective emissivity is on the order of $\epsilon_w=0.01$ \citep{SJ1972}. One exception is the area covered by the louvers of the thermal control system. On Pioneer 10 and 11, the effective emissivity of the closed louvers is approximately $\epsilon_l=0.04$ \citep{PC202}. As we do not have a separate value for New Horizons, we shall use the Pioneer figure. The emissivity changes as the louvers open partially, but the system also becomes highly non-Lambertian, as the angled louvers preferentially reflect internal heat in one particular direction. Due to the complexities of this scenario, we ignore the case when louvers are partially open, i.e., when the spacecraft is still near the Sun.

We assume that the spacecraft walls and the louver blades are Lambertian emitters. The side walls will not contribute to acceleration; heat radiated by these walls will, on average, be emitted in a direction perpendicular to the spin axis. Similarly, the wall facing the back of the HGA will not contribute to acceleration either; as it is facing the highly ($\sim$99\%) reflective backside of the HGA, very little heat will leave the spacecraft body in this direction. Therefore, the amount of heat contributing to acceleration can be calculated as the ratio of heat leaving through the bottom of the spacecraft (which is partially covered by louvers) to the total heat leaving through the bottom and sides:
{}
\begin{equation}
{P_{E_\mathrm{accel}}}=\frac{2}{3}\frac{\epsilon_w(A_b-A_l)+\epsilon_lA_l}{\epsilon_w(A_b-A_l+A_s)+\epsilon_lA_l} {P_E},
\end{equation}
where $P_E$ is the total electrically generated heat inside the spacecraft body, $P_{E_\mathrm{accel}}$ is the amount of electrically generated heat contributing to acceleration, $\epsilon$ is the emissivity, and $A$ is the area of a surface ($b$: bottom, $s$: sides). The factor of 2/3 is due to the assumption that the surfaces are Lambertian emitters.

No published data appears to exist on the actual size of the New Horizons louvers, but published ``artist's renderings'' appear to depict at least four louver assemblies with ten louver blades each. On Pioneer 10 and 11, a set of 30 louver blades covered an area of $\sim0.3\,\mathrm{m}^2$; therefore, we shall assume that the area of the louver system on New Horizons is $A_l\simeq 0.4\,\mathrm{m}^2$. The surface area of the rest of the craft body can be calculated using low-resolution, but dimensionally correct drawings, leading to the estimate
{}
\begin{equation}
P_{E_\mathrm{accel}}=0.46P_E=68.4\,\mathrm{W}.
\end{equation}

\subsection{Heat from the RTG}

Most of the power from the RTG is not converted to electricity but radiated away into space in the form of heat. The radiation pattern of the RTG is fore-aft symmetric, resulting in no net (average) acceleration force on a spinning spacecraft. Some of the heat, however, is reflected by the back of the HGA, resulting in an acceleration force.

A quick look at published drawings of the New Horizons spacecraft makes it clear that no parts of the spacecraft body stand in the way of thermal radiation emitted from the RTG in the direction of the back of the HGA. However, an approximately circular heat shield at the base of the RTG blocks some thermal radiation.

Any radiation that impacts the back of the HGA will transfer momentum to the spacecraft. Additionally, radiation that is reflected by the HGA transfers further momentum. To estimate the total amount of momentum transferred to the spacecraft this way, one needs to enumerate the following:

\begin{enumerate}[i)]
\item $P_{\mathrm{incident}}$, the spin-axis component of thermal radiation incident on the back of the HGA, as a function of the angle between the ray of radiation and the spin axis;
\item $P_{\mathrm{specular}}$, the spin-axis component of thermal radiation specularly reflected by the back of the HGA, as a function of the angle of incidence at the HGA and the angle of the HGA surface normal at the point of incidence relative to the spin axis;
\item $P_{\mathrm{diffuse}}$, the spin-axis component of thermal radiation diffusely reflected by the back of the HGA, as a function of the angle between the HGA surface normal at the point of incidence relative to the spin axis.
\end{enumerate}

To compute these quantities, we begin with the equation of heat transfer between two Lambertian surfaces \citep{Toth:2009se}:
{}
\begin{equation}
P_{1\rightarrow 2}=\displaystyle\int{\int{\frac{\cos{\theta_1}\cos{\theta_2}\mathcal{P}_1}{4\pi r^2}dA_2}dA_1},
\label{eq:heatxfer}
\end{equation}
where $A_1$ and $A_2$ represent the surface of the emitting and the absorbing body, respectively; $r$ is the distance between the surface elements $dA_1$ and $dA_2$; $\theta_1$ and $\theta_2$ are the angles between the line connecting the two surface elements and their respective normals; and $\mathcal{P}_1$ is the emitted power density (power per unit area) at the surface $A_1$.

We can eliminate one of the double integrals by noting that the RTG is approximately isothermal along its length, and it is approximately cylindrically symmetrical (indeed, the hexagonal fin arrangement means that one can substitute a cylindrical body of a diameter that is the mean of unity and $\cos{30^\circ}=\sqrt{3}/2$, introducing an error no larger than $\sim 7\%$; in actuality, as the fins are likely colder than the core of the RTG, the error will be even smaller.) Put together, these considerations help us reduce Eq.~(\ref{eq:heatxfer}) to
{}
\begin{equation}
P_{\mathrm{RTG}\rightarrow\mathrm{antenna}}=P_{\mathrm{RTG}}\displaystyle\int{\int{\frac{\sin{\beta}\cos{\theta}}{4\pi r^2}dL}dA},
\end{equation}
where $L$ represents the length of the RTG and $A$ represents the portion of the back of the RTG illuminated by that point of the RTG. As the RTG is assumed to be lengthwise isothermal, we could also move $P_{\mathrm{RTG}}$, denoting the RTG thermal power (total power minus power removed by the thermocouples in the form of electricity; $\sim$3360~W at the beginning of mission), outside the integration sign. $\beta$ now denotes the angle between the ray of radiation and the lengthwise RTG axis, and $\theta$ is the angle of incidence relative to the RTG surface normal.

To obtain the spin-axis component of this incident ray of radiation, we need to further multiply by $\sin{\beta}$:
{}
\begin{equation}
P_{\mathrm{incident}}=P_{\mathrm{RTG}}\displaystyle\int{\int{\frac{\sin^2{\beta}\cos{\theta}}{4\pi r^2}dL}dA}.
\end{equation}

For diffusely reflected radiation, we note that for a Lambertian reflector, momentum transferred will be in a direction perpendicular to the surface element, and it will be proportional to the incident radiation times $2\rho/3$, where $\rho$ denotes the reflectance. To calculate the spin-axis component of diffusely reflected radiation, we need to therefore compute
{}
\begin{equation}
P_{\mathrm{diffuse}}=\frac{2}{3}{\rho(1-\sigma)P_{\mathrm{RTG}}} \displaystyle\int{\int{\frac{\sin{\beta}\cos{\gamma}\cos{\theta}}{4\pi r^2}dL}dA},
\end{equation}
where $\gamma$ is the angle between the normal of the surface element $dA$ and the spin axis, and $\sigma$ is the ratio of specular vs. total reflected radiation.

Radiation that is specularly reflected will be emitted in accordance with the laws of geometric optics. Denoting the angle of specularly reflected radiation at surface element $dA$ by $\delta$, we can compute the spin axis component of specularly reflected radiation as
{}
\begin{equation}
P_{\mathrm{specular}}=\rho\sigma P_{\mathrm{RTG}}\displaystyle\int{\int{\frac{\sin{\beta}\cos{\delta}\cos{\theta}}{4\pi r^2}dL}dA}.
\end{equation}

The values of $\beta$, $\gamma$, $\delta$, and $\theta$, as well as the integration limits for the surface integrals (i.e., the boundaries of the HGA area illuminated by the RTG) can be computed using geometric considerations. An examination of the geometry also tells us that approximately 20\% of reflected radiation will be intercepted by the spacecraft body, and thus not contribute to thrust. The total amount of radiation that contributes to thrust in the direction of the spin axis can be summed as:
\begin{equation}
P_{\mathrm{thrust}}=P_{\mathrm{incident}}+0.8(P_{\mathrm{diffuse}}+P_{\mathrm{specular}}).
\end{equation}

Numerical evaluation of these integrals yields
{}
\begin{equation}
P_{\mathrm{thrust}}=0.15P_{\mathrm{RTG}},
\end{equation}
as the amount of thermal power, as a function of total RTG power $P_{\mathrm{RTG}}$, that will contribute to acceleration along the spin axis. At the beginning of mission, this translates into approximately 500~W of power contributing to acceleration.

While the approximations used in this section are clearly no substitute for the evaluation of a detailed finite element model of the spacecraft, the result indicates that anisotropic thermal radiation is a potentially significant source of acceleration for New Horizons during its long interplanetary cruise.

\subsection{Acceleration due to emitted heat}

The relationship between the momentum and energy of electromagnetic radiation is well known: $p=E/c$. To calculate the acceleration due to collimated electromagnetic radiation of power $P$ emitted by a body of mass $m$, one can use the formula $a=P/(mc)$. The nominal mass at launch of the New Horizons spacecraft is about 465~kg \citep{NHFEIS1}. Thus, we can calculate an anomalous acceleration for the New Horizons spacecraft due to thermal radiation from the RTG and electrical equipment as
{}
\begin{equation}
a_{\rm NH}=41\times 10^{-10}\,\mathrm{m}/\mathrm{s}^2\simeq4.7\,a_P.
\end{equation}

When the spacecraft's 15~W transmitter is operating, the power of the radio beam (emitted in the direction opposite to the direction of thermal effects), which translates into an acceleration of $1\times 10^{-10}\,\mathrm{m}/\mathrm{s}^2$, would have to be subtracted from our result.

The availability of the entire history of the Pioneer spacecraft makes it possible for us to calculate acceleration due to on-board forces to a significantly greater precision, and as a function of spacecraft parameters that evolve with time. The contribution of on-board forces to acceleration can be significant. Therefore, preserving and making available raw (engineering) telemetry of New Horizons to researchers is of great importance if accurate orbit determination is desired, and especially if conclusions are to be derived from any discrepancies between computed and observed orbits.

\section{Conclusions}

By 2009, the existence of the Pioneer anomaly is no longer in doubt. Our continuing effort to process and analyze Pioneer radio-metric and telemetry data is part of a broader strategy \citep{JPL2005,MDR2005}.

Based on the information provided by the telemetry, we were able to develop a high accuracy thermal, electrical, and dynamical model of the Pioneer spacecraft. This model is used to investigate the anomalous acceleration and especially to study the contribution from the on-board thermal environment to the anomaly.

The available thermal model for the Pioneer spacecraft accounts for all heat radiation produced by the spacecraft. In fact, we use telemetry information to accurately estimate the amount of heat produced by the spacecrafts' major components. We are in the process of evaluating the amount of heat radiated in various directions.

This entails, on the one hand, an analysis of all available radio-metric data, to characterize the anomalous acceleration beyond the periods that were examined in previous studies. Telemetry, on the other hand, enables us to reconstruct a thermal, electrical, and propulsion system profile of the spacecraft. Soon, we should be able to estimate effects on the motion of the spacecraft due to on-board systematic acceleration sources, expressed as a function of telemetry readings. This provides a new and unique way to refine orbital predictions and may also lead to an unambiguous determination of the origin of the Pioneer anomaly.

Concluding, we mention that before Pioneer 10 and 11, Newtonian gravity had never been measured---and was therefore never confirmed---with great precision over great distances. The unique ``built-in'' navigation capabilities of Pioneer 10 and 11 allowed them to reach the levels of $\sim10^{-10}$ m/s$^2$ in acceleration sensitivity. Such an exceptional sensitivity means that Pioneer 10 and 11 represent the largest-scale experiment to test the gravitational inverse square law ever conducted.  However, the experiment failed to confirm the validity of this fundamental law of Newtonian gravity in the outer regions of the solar system. One can demonstrate, beyond 15 AU the difference between the predictions of Newton and Einstein are negligible. So, at the moment, two forces seem to be at play in deep space: Newton's laws of gravity and the mysterious Pioneer anomaly. Until the anomaly is thoroughly accounted for by natural causes, and can therefore be eliminated from consideration, the validity of Newton's laws in the outer solar system will remain unconfirmed. This fact justifies the importance of the investigation of the nature of the Pioneer anomaly.

\begin{acknowledgements}
This work was partially performed at the International Space Science Institute (ISSI), Bern, Switzerland, when both of us visited ISSI as part of an International Team program. In this respect we thank Roger M. Bonnet, Vittorio Manno, Brigitte Schutte and Saliba F. Saliba of ISSI for their hospitality and support. We especially thank The Planetary Society for support and, in particular, Louis D. Freidman, Charlene M. Anderson, and Bruce Betts for their interest, stimulating conversations and encouragement.

The work of SGT was carried out at the Jet Propulsion Laboratory, California Institute of Technology, under a contract with the National Aeronautics and Space Administration.
\end{acknowledgements}

\end{document}